# Inference of ventricular activation properties from non-invasive electrocardiography

Julia Camps[1*], Brodie Lawson[2,3], Christopher Drovandi[2,3], Ana Minchole[1], Zhinuo Jenny Wang[1], Vicente Grau[4], Kevin Burrage[1,2], Blanca Rodriguez[1*]

[1]Department of Computer Science, University of Oxford, Oxford, United Kingdom.

[2]Australian Research Council Centre of Excellence for Mathematical and Statistical Frontiers (ACEMS), Queensland University of Technology (QUT), Brisbane, Australia.

[3]QUT Centre for Data Science (CDS), Queensland University of Technology, Brisbane, Australia.

[4]Institute of Biomedical Engineering (IBME), University of Oxford, Oxford, United Kingdom.

*Corresponding author: {julia.camps, blanca}@cs.ox.ac.uk

Abstract

The realisation of precision cardiology requires novel techniques for the non-invasive characterisation of individual patients' cardiac function to inform therapeutic and diagnostic decision-making. The electrocardiogram (ECG) is the most widely used clinical tool for cardiac diagnosis. Its interpretation is, however, confounded by functional and anatomical variability in heart and torso. In this study, we develop new computational techniques to estimate key ventricular activation properties for individual subjects by exploiting the synergy between non-invasive electrocardiography and image-based torso-biventricular modelling and simulation. More precisely, we present an efficient sequential Monte Carlo approximate Bayesian computation-based inference method, integrated with Eikonal simulations and torso-biventricular models constructed based on clinical cardiac magnetic resonance (CMR) imaging. The method also includes a novel strategy to treat combined continuous (conduction speeds) and discrete (earliest activation sites) parameter spaces, and an efficient dynamic time warping-based ECG comparison algorithm. We demonstrate results from our inference method on a cohort of twenty virtual subjects with cardiac volumes ranging from 74 $cm^3$ to 171 $cm^3$ and considering low versus high resolution for the endocardial discretisation (which determines possible locations of the earliest activation sites). Results show that our method can successfully infer the ventricular activation properties from non-invasive data, with higher accuracy for earliest activation sites, endocardial speed, and sheet (transmural) speed in sinus rhythm, rather than the fibre or sheet-normal speeds.

**Keywords:** Earliest activation sites; Parameter inference; Electrocardiogram; Cardiac Magnetic Resonance.

## 1 Introduction

Cardiovascular diseases account for 31% of deaths globally, according to the World Health Organisation (2016). Cardiac disease increases the risk of sudden and premature death through alterations in cardiac electrophysiology and tissue structure, which are known to promote lethal arrhythmias and mechanical dysfunction.

Clinically, the electrocardiogram (ECG) is the most widely used modality for diagnosis. The information that can be extracted from the ECG is, however, confounded by anatomical and functional variability in the human population. For example, variability in the heart location and orientation within the torso affects the morphology of the QRS complex (Mincholé et al. 2019). Non-invasive imaging, through ultrasound, computerised tomography or cardiac magnetic resonance (CMR), is also used clinically to provide further information on cardiac anatomy, structure and mechanical function. Novel techniques are needed to fully exploit the synergy that can be obtained by combining ECG and non-invasive clinical modalities such as CMR.

Recent technology has shown the power of CMR-based modelling and simulation of the ECG to aid in the identification and interpretation of patients' phenotypic variability. Patient-specific anatomical models of the cardiac ventricles embedded in the torso are generated from CMR and used to simulate the ECG using electrophysiological models of propagations through the ventricles. Comparison of simulated and clinical ECGs under the different tested conditions allows the identification of critical electrophysiological and structural factors that help to explain each ECG phenotype.

In this study, we aim to investigate new computational techniques for the quantification of subject-specific ventricular properties from non-invasive electrocardiographic data using CMR-based modelling and simulation. This study presents an efficient inference method combined with fast ECG simulations using CMR-based torso-biventricular models to determine the accuracy in the estimation of ventricular activation properties from the 12-lead ECG or activation time maps (as obtained through electrocardiographic imaging). We conduct the simultaneous inference of endocardial and myocardial conduction speeds, and the location of earliest sites of activation in the endocardium (also called root nodes), as these properties determine the activation sequence in the ventricles. To address the challenges associated with inferring root node locations and speeds simultaneously, we implement a novel inference method based on the sequential Monte Carlo approximate Bayesian computation (SMC-ABC) algorithm (Sisson, Fan, and Tanaka 2007; Drovandi and Pettitt 2011). From our experiments, we quantified the accuracy of recovering the activation properties from two data modalities (12-lead ECG and epicardial activation maps); and twenty torso-biventricular models with ventricular volumes and electrophysiological properties. Our analysis will aid future works addressing this, previously unexplored, subject-specific calibration problem in both ventricles simultaneously.

## 2   Materials and Methods

Fig. 1 presents a diagram of our proposed approach, including input, output, and the iterative process, to infer the human ventricular activation properties from 12-lead ECG recordings. The input data (grey box, top) include 'given data' (torso-biventricular mesh, and fibre orientations) as well as the 'target data' (electrocardiographic recording). The output of the inference process (grey box, bottom) are the sets of parameter values (also called particles) from the population of models, as in Britton et al. (2013), for endocardial, fibre, sheet (transmural), and sheet-normal conduction speeds, and root node locations (also called earliest sites of activation).

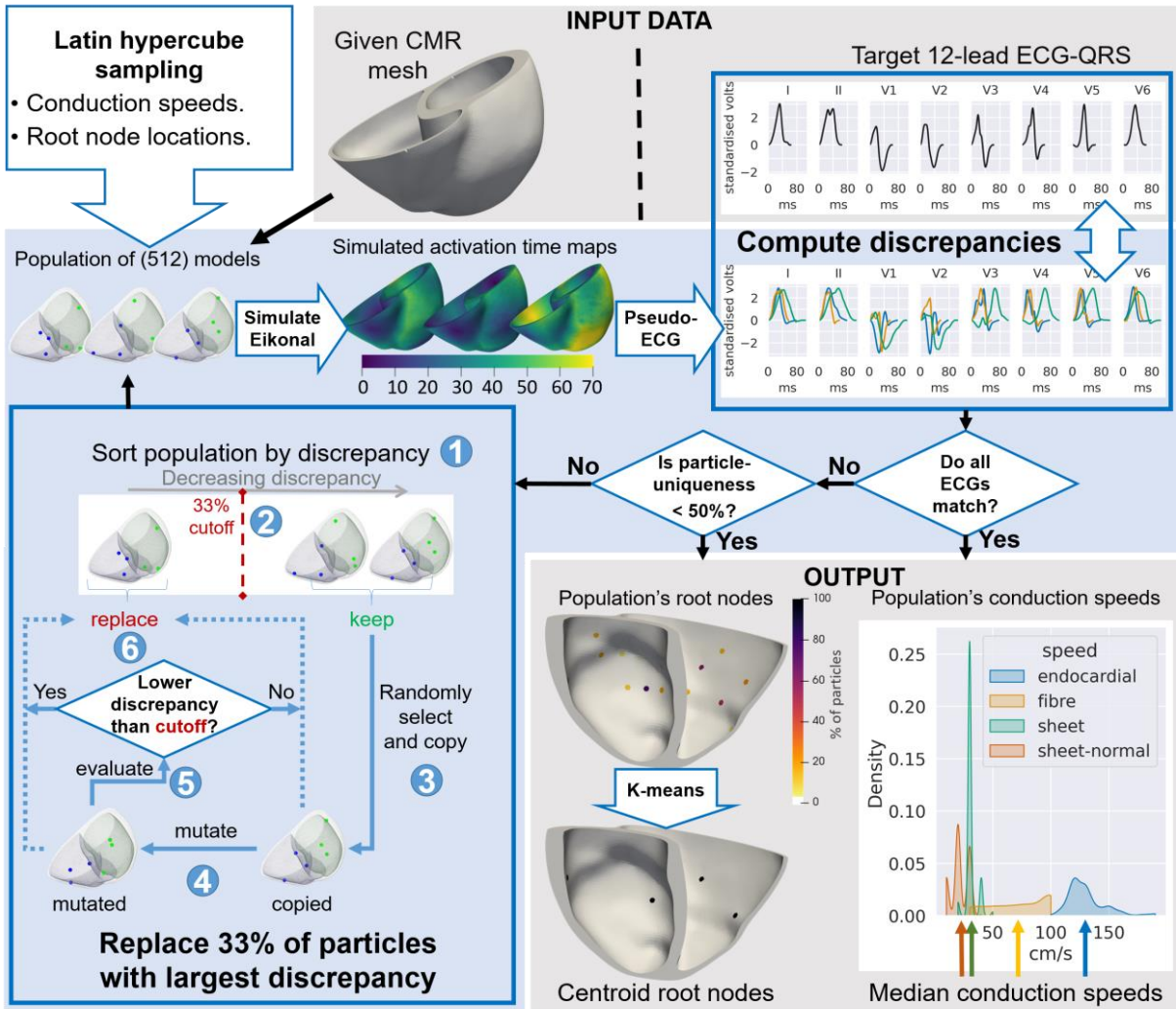

*Fig. 1. Proposed inference method. Diagram of our inference pipeline, and its subprocesses, to recover the activation properties from CMR and 12-lead ECG data. The process starts with the Latin hypercube sampling generating a population of 512 particles, which combined with the 'Given CMR mesh' becomes a population of 512 electrophysiological (Eikonal) models. From there, the iterative process (light blue shadowed area) follows in the direction of the arrows until either of the two stopping criteria is fulfilled. The 'Input' and 'Output' data areas are shaded in grey, on the top and bottom of the figure, respectively.*

The process starts with the creation of the population of 512 particles (parameter sets for conduction speeds and root nodes) by the Latin hypercube sampling (Iman and Conover 1980) and the 'Given CMR mesh' (left part of the Input data, top grey box) to create a population of 512 biventricular electrophysiological models. The iterative process followed for the inference (blue shaded area) includes five steps: 1. simulate the new population of particles using the CMR-based models; 2. generate the simulated ECGs; 3. compute the discrepancies with the target ECG; 4. evaluate the stopping condition; and, 5. modify the population of particles using the feedback from the discrepancies. For this work, we simulated the electrophysiology using CMR-based Eikonal models with rule-based fibre orientations (Streeter et al. 1969), orthotropic conduction speeds, and isotropic endocardial conduction layer to emulate the Purkinje network (Cardone-Noott et al. 2016).

In Step 1, activation time maps are simulated for each parameter set in the population using the Eikonal equation (Appendix A.2), which was solved with Dijkstra's algorithm (Dijkstra 1959). In Step 2, the 12-lead ECG-QRS is computed from each activation time map using the pseudo-ECG algorithm (Appendix A.4) (Gima and Rudy 2002). Then, Step 3 computes the discrepancy (distance metric) between the simulated and target 12-lead ECG-QRS using an extension (Section 2.4) of the dynamic time warping (DTW) (Velichko and Zagoruyko 1970) algorithm. Step 4 checks the two stopping criteria: 1. all simulated 12-lead ECGs 'match' the target ones, namely, all particles in the population have discrepancies smaller than the desired positive tolerance (hyperparameter) – a discrepancy of zero is usually impossible to attain due to differences between data sources; 2. less than 50% of the particles in the population are 'unique'. If the population does not fulfil either criterion, the iterative process continues, and all particles are sent to Step 5.

Step 5 replaces particles with the highest discrepancy (computed in Step 3) values (poor regions of the parameter space) with those with new parameter sets from promising regions of the parameter space (Section 2.3 and Appendix A.5). More precisely, Step 5 can be seen as the sequence of the following six substeps: 1. sort the population of particles according to their discrepancy values; 2. label the one-third of particles in the population with the highest discrepancy as 'to replace' and the remaining ones as 'to keep'; 3. for each 'to replace' particle randomly select and copy one particle from the group 'to keep'; 4. mutate the copied particles; 5. compute the discrepancy of each mutated particle; and, 6. If the discrepancy of the mutated particle is lower than the one third cutoff value used to split the two groups, replace the corresponding 'to replace' particle with the mutated one, otherwise, use the copied particle for the replacement.

After Step 5, the iterative process restarts from Step 1 with the modified population of models, and the inference process carries on until one of the stopping criteria in Step 4 is met. The same approach (without the pseudo-ECG) solves the inference guided by epicardial activation time map data.

The 'output' area of Fig. 1 (bottom right – grey shade) also illustrates the post-processing of the parameter values in the resulting population. We combined all values of each parameter in the population to calculate a single solution parameter set for each inference experiment. The strategy to combine our particles takes the median value for each conduction speed across all particles in the population. The solution set of root nodes locations are the centroids of clusters found by applying k-means clustering (initialised at the most frequently occurring configuration for the root node parameters in the population) to this same resulting population.

## 2.1 Virtual Subjects and Clinical Datasets

Our validation dataset included the epicardial activation maps and 12-lead ECGs generated through bidomain (Appendix A.1) simulations from a cohort of 20 healthy virtual subjects. These were obtained as described in Mincholé et al. (2019) using four torso-biventricular meshes constructed from the CMRs of four healthy people, five conduction speed scenarios (Appendix A.3), as outlined in Table 1.

| Conduction configuration | Endocardial speed | Fibre speed | Sheet speed | Sheet-normal speed |
|---|---|---|---|---|
| Normal speeds | 150 cm/s | 50 cm/s | 32 cm/s | 29 cm/s |
| Slow endocardial speed | 120 cm/s | 50 cm/s | 32 cm/s | 29 cm/s |
| Fast endocardial speed | 179 cm/s | 50 cm/s | 32 cm/s | 29 cm/s |
| Fast endocardial and myocardial speeds | 179 cm/s | 88 cm/s | 49 cm/s | 45 cm/s |
| Slow endocardial and fast myocardial speeds | 120 cm/s | 88 cm/s | 49 cm/s | 45 cm/s |

*Table 1. The five conduction-speed configurations considered for this study. These Eikonal speed values were selected in order to match the bidomain (Mincholé et al. 2019) conductivities (Appendix A.3).*

More precisely, we considered four torso-biventricular CMR-based meshes with a variable torso and ventricular volumes and orientations to explore the effects of anatomical variability. These four meshes had the following torso-volumes: 23 dm$^3$, 27 dm$^3$, 54 dm$^3$, and 44 dm$^3$ for Mesh-1, Mesh-2, Mesh-3, and Mesh-4, respectively. The biventricular volumes for these meshes were 74 cm$^3$, 76 cm$^3$, 107 cm$^3$, and 171 cm$^3$, respectively.

## 2.2  Parameter space exploration

Our approach aims to infer the root node locations and the four conduction speeds (i.e. endocardial, fibre, sheet, and sheet-normal speeds) that enable the model to reproduce a subject's ECG recording (or epicardial activation map). Two sets of possible root node locations were considered, namely, low and high resolution of the root node discretisation. In the low resolution of the root node discretisation, the possible root node locations were preselected as the centres of each endocardial section following the American Heart Association's segmentation guidelines (Cerqueira Manuel D. et al. 2002). This selection strategy led to 39 candidate locations (17 in the left and 22 in the right endocardium). A high resolution of the root node discretisation was also considered with 102 potential root nodes (62 and 40 candidate nodes for the left and right endocardium, respectively) by adding additional potential root nodes to fill the gaps in the low-resolution discretisation. The candidate root node locations were considered as a binary parameter, with the inference yielding 'in use' or 'not in use' to obtain a good match between simulated and target electrocardiographic signal.

The conduction speeds were assigned to be within predefined physiological ranges (Durrer et al. 1970). More precisely, the endocardial and myocardial speeds were bounded within the ranges [100, 200] and [10, 100] cm/s, respectively. Furthermore, we constrained the fibre speed to be larger than the sheet speed, and sheet speed to be larger than the sheet-normal speed. Finally, we set a uniform initial distribution for all conduction speeds for the Latin hypercube sampling. For the root nodes, we set an initial distribution on the number of root nodes to be selected based on the findings from Cardone-Noott et al. (2016) for a healthy and realistic ECG generation. More precisely, we set this initial distribution as a truncated normal distribution centred at seven locations with a standard deviation of two over the range [2, 14] root nodes. We considered no ventricle-specific or location preferences.

## 2.3  Parameter inference with SMC-ABC method

The combination of conduction speeds and root nodes creates an inference problem with continuous and discrete mixed-type parameter space that challenges many parameter inference algorithms. We propose the application of an SMC-ABC (Appendix A.5) (Sisson, Fan, and Tanaka 2007; Drovandi and Pettitt 2011) algorithm that we specifically designed to explore our mixed-type parameter space efficiently. In a nutshell, SMC-ABC defines its intermediate distributions (SMC) as approximate posteriors with a series of decreasing cutoff discrepancy values (ABC). In our context, SMC-ABC serves as a particle-based optimisation approach that solves a sequence of easier optimisation problems where each informs the next.

In generating new proposals for SMC-ABC's mutation steps (Step 5 in Fig. 1), we ignore any dependencies between the discrete (root node locations) and continuous (conduction speeds) spaces. For the continuous space, we use the typical approach of random Gaussian jumps informed by the current covariance of the particles. For the discrete root node space, however, we propose a novel strategy to generate proposals that are independent of a particle's current parameter values but informed by the whole population of particles. Independent proposal distributions are asymmetric, but can significantly accelerate SMC approaches by generating proposals where large jumps are more likely to be accepted (South, Pettitt, and Drovandi 2019).

More precisely, we first determine the number of root node locations for a new particle. To choose the number of root nodes, 80% of the time, we take the number of root nodes from a random particle in the population. The remaining 20% of the time, we sample the number of root nodes from a normal distribution with a mean of seven and a standard deviation of two (to prevent premature convergence). Next, we place a Dirichlet (i.e. multivariate Beta) prior on the propensities that a candidate root node location is 'in use'. We treat the frequency of a candidate location being 'in use' in our population as a sample from a multinomial distribution. This choice of treatment allows us to calculate these propensities from the current set of particles since the Dirichlet distribution is the conjugate prior for a multinomial likelihood. Note that, due to the properties of conjugate priors, our posterior on propensities for the candidate root node locations will also be Dirichlet distributed. Given the number of root nodes to be 'in use', we select them one at a time by first sampling from the Dirichlet posterior implied by the 'relevant' particles in the population to obtain a set of propensities. Then we sample from the (single-trial) multinomial distribution with these propensities.

A particle is 'relevant' if it has the same number of root nodes as the particle being created, and also the same root nodes locations 'in use' as those selected so far during this process. Note that, the population of relevant particles that informs the selection of the next candidate root node location becomes smaller after each selection. Consequently, the first few selections of root nodes will tend to represent the existing patterns in the population of particles, while the last selections should serve to explore new candidate root node locations. Importantly, by taking this approach and only using matching particles to update the propensities each time a candidate root node is selected, we capture interdependencies between the locations that are 'in use' across the particle population.

## 2.4 Discrepancy calculation

For the epicardial activation map domain, we use the mean over nodes of the absolute difference between two epicardial activation maps as the discrepancy metric. For the ECG domain, we propose a novel extension of the DTW algorithm as a discrepancy metric.

Measuring differences between ECG recordings is challenging, especially in a way that captures the resemblance of the source activation properties. DTW is a speed-invariant dynamic-programming algorithm for measuring differences between sequences. Hence, DTW can compare signals of different length, such as ECGs by stretching and shrinking them in the time axis. Classic DTW allows ECGs to be compared using non-physiological warping (Luzianin and Krause 2016). A window warping-constraint, as in Sakoe and Chiba (1978), can restrict the maximum warping to avoid non-physiological transformations. However, a static window-constraint DTW algorithm would be computationally inefficient to explore the parameter space. For example, a static window-constraint cannot distinguish between two versions of the same signal with different stretching/shrinking when both are within the window. Moreover, deciding on the size of the window is complicated as we want a wide window to identify matching morphologies and also a short window to enforce the matching of the duration of the target QRS.

We assume that in the healthy setting, the root node locations are responsible for the QRS's morphology and that the conduction speeds define QRS's width (Cardone-Noott et al. 2016). This assumption allows us to separate our optimisation process into two distinct phases: 1) recovering the root nodes from the morphology of the 12-lead ECG-QRS while preserving all speed values; and, once all particles produce acceptable morphologies, 2) iteratively narrow down the population to the particles with the conduction speeds that generate the most similar QRS width to the target recording.

In other words, our DTW algorithm should be more forgiving with the amount of warping allowed initially when all the particles produce drastically different QRS's morphologies and widths. This initial gentleness of the metric will ensure that the correct morphologies are identified and preserved from the start, which is when the distribution of particles is most diffuse. Then, the metric should become stricter on the amount of warping allowed towards the end of the iterative process.

Our DTW-based discrepancy metric implements a shrinking-window warping-constraint that gradually decreases in size throughout the iterative process preserving the particles with the best morphology in the QRS along the way. Therefore, by the end of the iterative process, the size of the shrinking-window is almost zero, and, thus, the DTW-based discrepancy resembles the absolute difference between signals.

However, since the morphology of the QRS is determined by the root node locations, all the particles in the population with the same root nodes, regardless of their speeds, will receive the same discrepancy value. Namely, the convergence of the speeds will always require the shrinking-window to close until almost no warping is allowed. Therefore, we defined a small warping penalty to improve the efficiency of our method by aiding the inference of the speeds in the case of negligible morphological variability among the ECGs generated from the population.

## 2.5 Simulation protocol

For each virtual subject, we define four inference experiment protocols, namely, two resolutions of the root node discretisation (high and low), and two target data domains (epicardial activation time map and 12-lead ECG-QRS). Moreover, we repeated all experiment protocols five times to ensure consistency and reproducibility of our results. The total number of inference experiments was 400 (20 virtual subjects 'times' 4 protocols 'times' 5 repetitions), which we run one at a time on a virtual machine with 18 2nd generation Intel Xeon Scalable Processors.

## 2.6 Hyperparameter calibration

Due to the lack of previous work on this problem, we calibrated the algorithms presented in this study using physiological knowledge, and adjusting the remaining variables so that our method would work in one of the 20 virtual subjects (Mesh-4 with 'Normal speeds' from Table 1) using the low resolution for the root node discretisation. Therefore, all our results share the same calibration values regardless of the virtual subject, the target data-type, or the resolution of the root node discretisation.

## 2.7 Error metrics

For this study, we adopted the following three error metrics: the epicardial activation time map prediction error, the 12-lead ECG prediction error, and the conduction speed inference error. Our error metrics allowed the aggregation of the results sharing the same anatomy. More precisely, we computed the error in the prediction of the target data as the mean of the per point relative activation time percentage errors, $mean(100 * \frac{P_i - T_i}{T_i})$, where $P_i$ and $T_i$ are the activation times for the epicardial node $i$ from the Eikonal-predicted and bidomain target activation time maps, respectively.

Similarly, for the ECG-guided inference, we computed the prediction percentage error $mean(100 * \frac{P_i - T_i}{\text{range}(T)})$, where $P_i$ and $T_i$ amplitude values for time $i$ in the standardised ECGs from the Eikonal-predicted and bidomain-target ECGs, respectively. We normalise the ECG error by the range (maximum minus minimum) of the standardised target QRS complex, $\text{range}(T)$.

The inference errors for the conduction speeds were computed as a percentage of the ground-truth value, namely, $100 * (S' - S)/S$, where $S'$ is the inferred endocardial speed, and $S$ is the endocardial speed employed in the bidomain (the ground truth value for the endocardial speed). Note that, the root node results could be directly visualised on the same geometry.

## 3 Results

### 3.1 Prediction of the activation time maps and ECGs target data

Fig. 2 illustrates the ability of our inference method to replicate the epicardial activation maps (top) and ECGs (bottom) for one of the virtual subjects. Results are shown using the low and high resolution for the root node discretisation. Moreover, the ECG part of the figure accounts for the simulated ECGs from all five repetitions of the experiment to assess the method's consistency, namely, 2560 QRSs (512 particles in one experiment 'times' 5 repetitions).

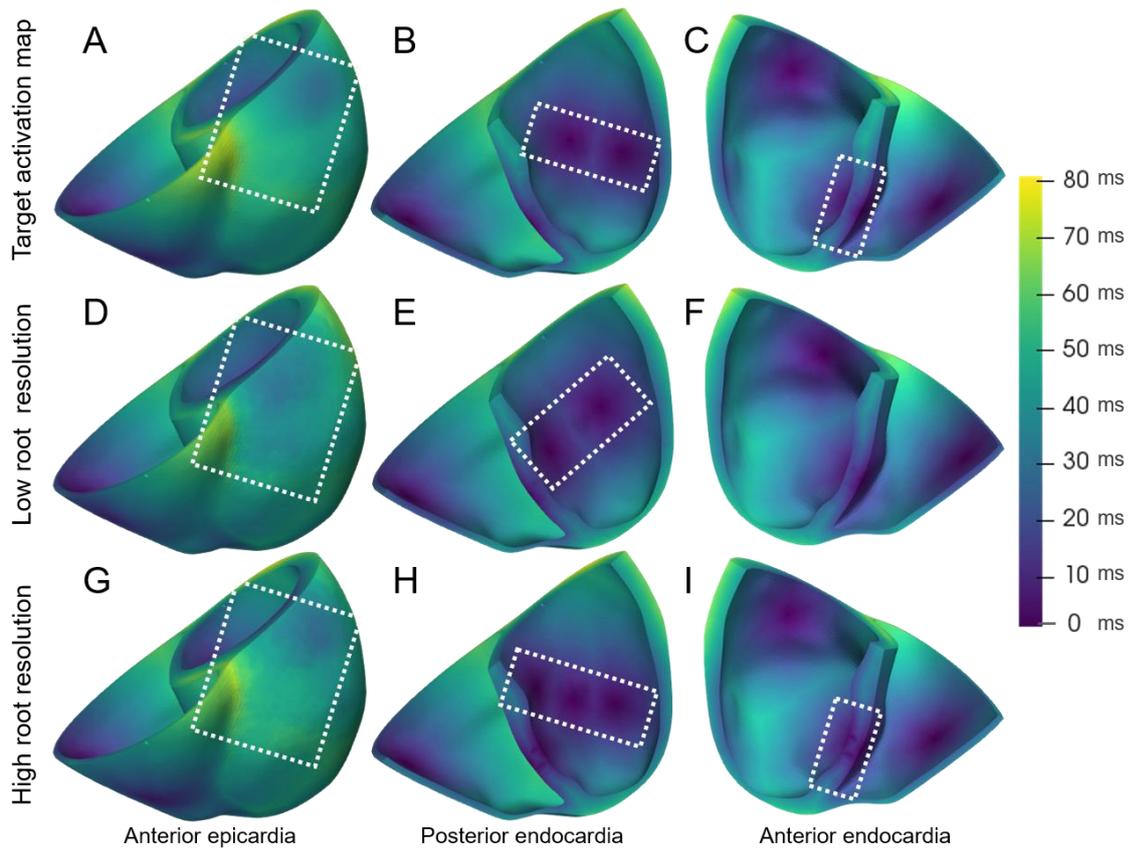
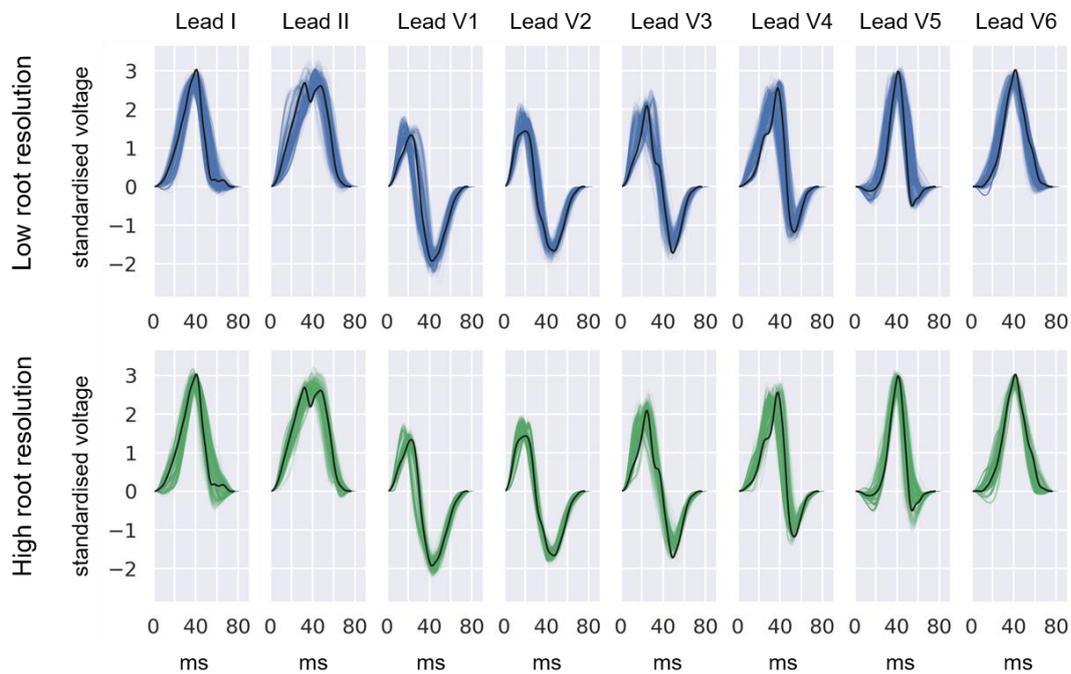

Fig. 2. Example of predicted activation map and ECGs from the inference. This figure illustrates three activation maps (one per row) that correspond, from top to bottom, to the target data, the simulation from the inference using the low resolution for the root node discretisation, and the simulation from the high resolution. The lower half of the figure shows the target QRSs (black) for the inference from ECG data, and the simulated QRSs from the resulting populations when using the low resolution for the root node discretisation (blue – top row) and the high resolution (green – bottom row). The ECG plots include the simulations from five repetitions of the same experiment to demonstrate consistency. These results corresponded to the inference process on the virtual subject with Mesh-4 and 'Slow endocardial speed' properties (see Table 1).

Fig. 2 shows that our simulated activation time maps and ECGs, using the virtual models, reproduced the input data consistently for both resolutions of the root node discretisation, and electrophysiological data modalities. The first column of the figure suggests that the high resolution for the root node discretisation (Fig. 2.G) was able to replicate better the input data (Fig. 2.A) than the low resolution (Fig. 2.D). The second column of the figure demonstrates that the high-resolution strategy (Fig. 2.H) overestimated the number of root nodes.

In contrast, the low resolution (Fig. 2.E) was able to correctly identify the number of root nodes, despite locating them slightly less accurately. The last column of the figure shows that the activation times close to the septum (Fig. 2.I compared to Fig. 2.C) were less accurately recovered than for the other regions of the heart.

Overall, our inference method was able to accurately reproduce the input epicardial activation maps with 6.8% and 12.2% median prediction errors (Section 2.7) when adopting the high and low resolutions for the root node discretisation, respectively. Similarly, for the ECG-based inference experiments, the median prediction errors (Section 2.7) were 7% and 14% for the high and low resolutions, respectively.

### 3.2 Root node inference from epicardial activation maps

The root node inference results are grouped by anatomy since the ground truth locations for all virtual subjects (with different speeds) sharing the same anatomy were identical. Fig. 3 illustrates the inferred root nodes locations from all experiments for the five virtual subjects with Mesh-4 (see Appendix A.6.1 for Mesh-1, Mesh-2, and Mesh-3).

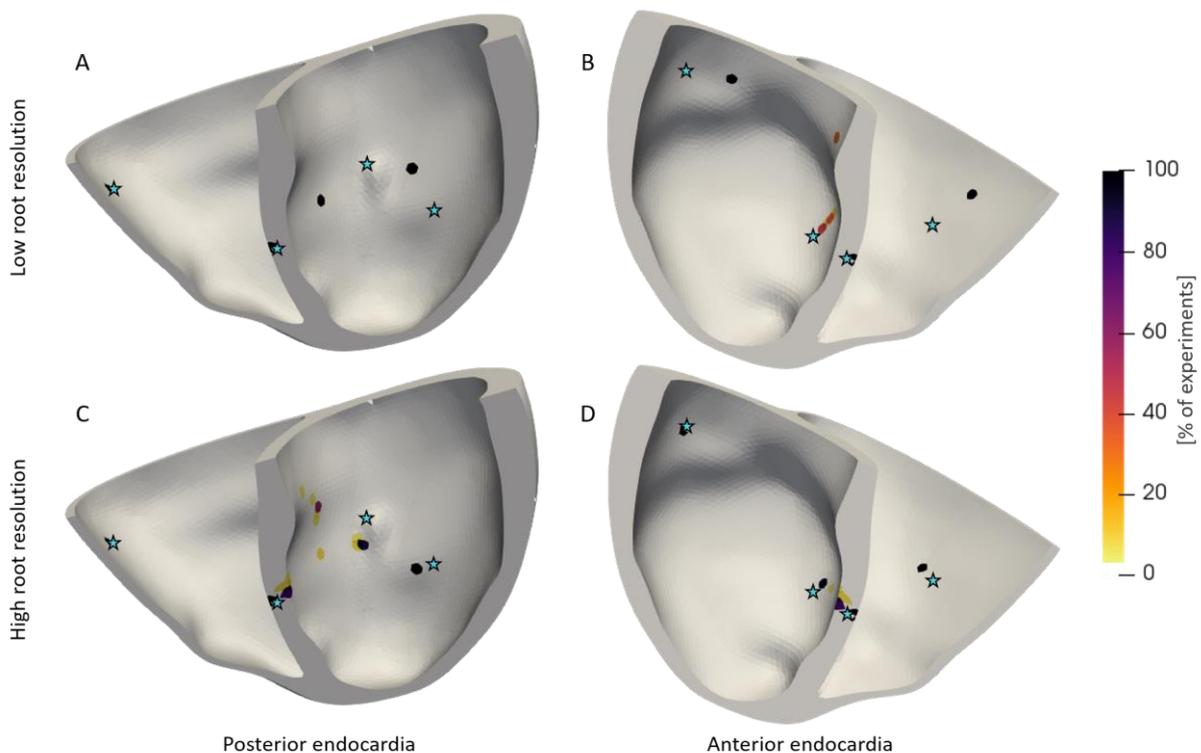

*Fig. 3. Root nodes inferred from activation maps on Mesh-4. This figure shows the root node locations inferred from the experiments with epicardial activation time maps as the target and with this torso-biventricular mesh. Subfigures 3.A and 3.B correspond to the predictions from using the low resolution for the root node discretisation; 3.C and 3.D correspond to the high resolution for the root node discretisation. The colourmap units are the percentage of the 25 inference experiments (5 repetitions 'times' 5 virtual subjects) that predicted a specific location. The star-icons denote the ground-truth locations.*

Fig. 3, as well as Fig. A.2, Fig. A.3, and Fig. A.4 in Appendix A.6.1, show that the recovery of the root nodes from epicardial activation maps was successful. The root node recovery accuracy is better for root nodes in non-septal areas since the root nodes in the septum have little influence on the epicardial activation time map. Moreover, the high-resolution discretisation achieved higher recovery accuracies compared to the low resolution, whereas these recovery accuracies were unaffected by the conduction speed configuration in each virtual subject.

### 3.3 Speed inference from epicardial activation maps

Fig. 4 illustrates the histogram of the percentage error in the estimation of each conduction speed from the experiments with epicardial activation maps as target and the high resolution for the root node discretisation. Each subfigure corresponds to the distribution of the inference percentage errors for one specific speed (Section 2.7). More precisely, subfigures 4.A, 4.B, 4.C, and 4.D correspond to the errors from the endocardial, the fibre, the sheet, and the sheet-normal speeds, respectively. In all histograms, the results are grouped by the anatomy of the virtual subjects. This visualisation strategy has enabled us to discuss the effect of anatomical differences in the subjects' hearts on the performance of our approach.

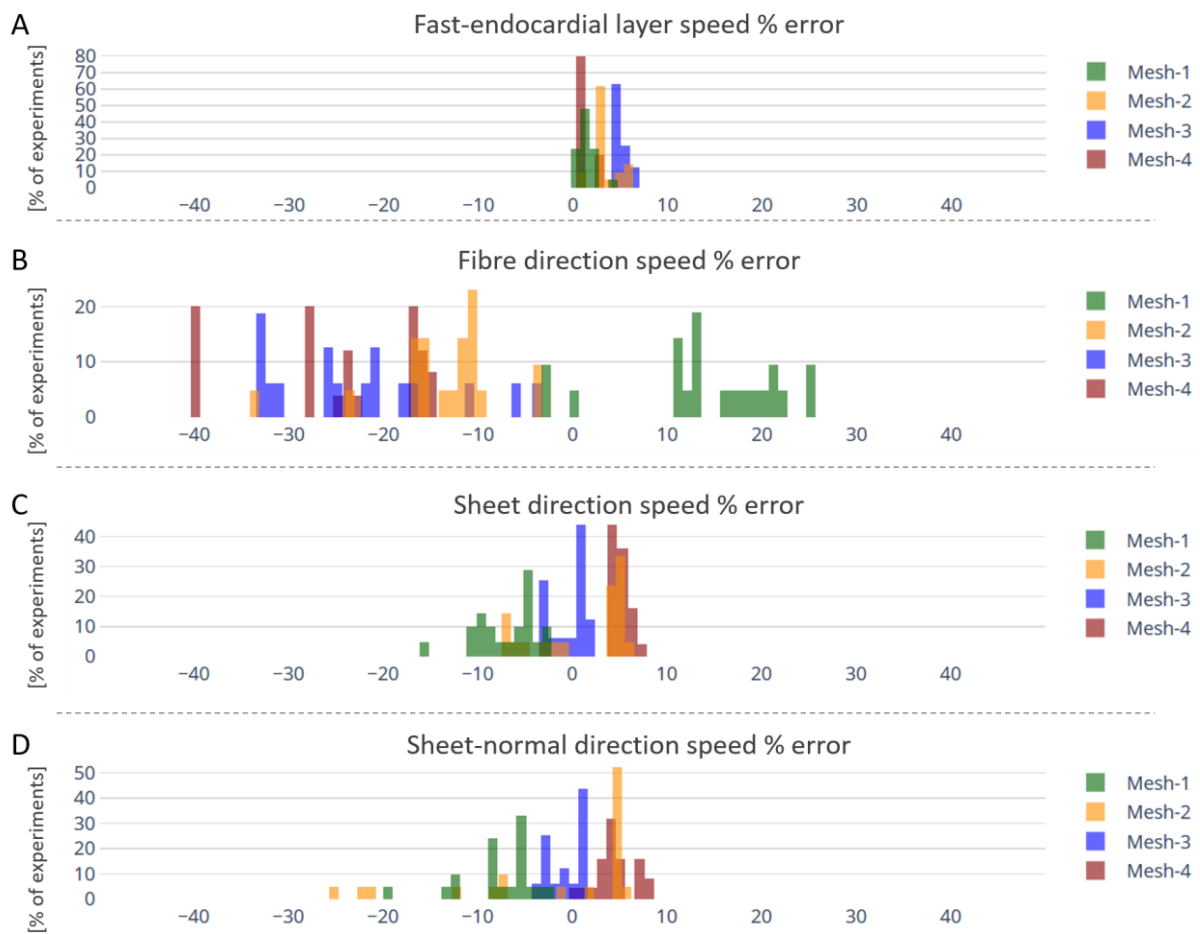

*Fig. 4. Error in the conduction speeds inference from activation maps and high resolution for the root node discretisation. The figure shows the inference errors as the percentage over the ground truth conduction speeds (x-axis) represented as histograms grouping the results by anatomy. The y-axis represents the percentage of experiments with a given inference error from the 25 experiments per anatomy (5 speed-configurations 'times' 5 repetitions). Subfigures 4.A, 4.B, 4.C, and 4.D correspond to the errors from the endocardial, the fibre, the sheet, and the sheet-normal speeds, respectively.*

Fig. 4 reports inference errors (Section 2.7) with absolute values less than 7%, 40%, 13%, and 26% for the endocardial, fibre, sheet, and sheet-normal speeds, respectively, for all the featured experiments. The median of these absolute percentage errors were 2.3%, 16.8%, 4.7%, and 4.6%, for the mentioned speeds, respectively. The recovery errors for the fibre speed were consistently higher than for the other speeds in all experiments featured in this figure.

Analogously to Fig. 4, Fig. 5 presents the inference percentage error for each speed from the experiments that used epicardial activation time maps as the target and low resolution for the root node discretisation.

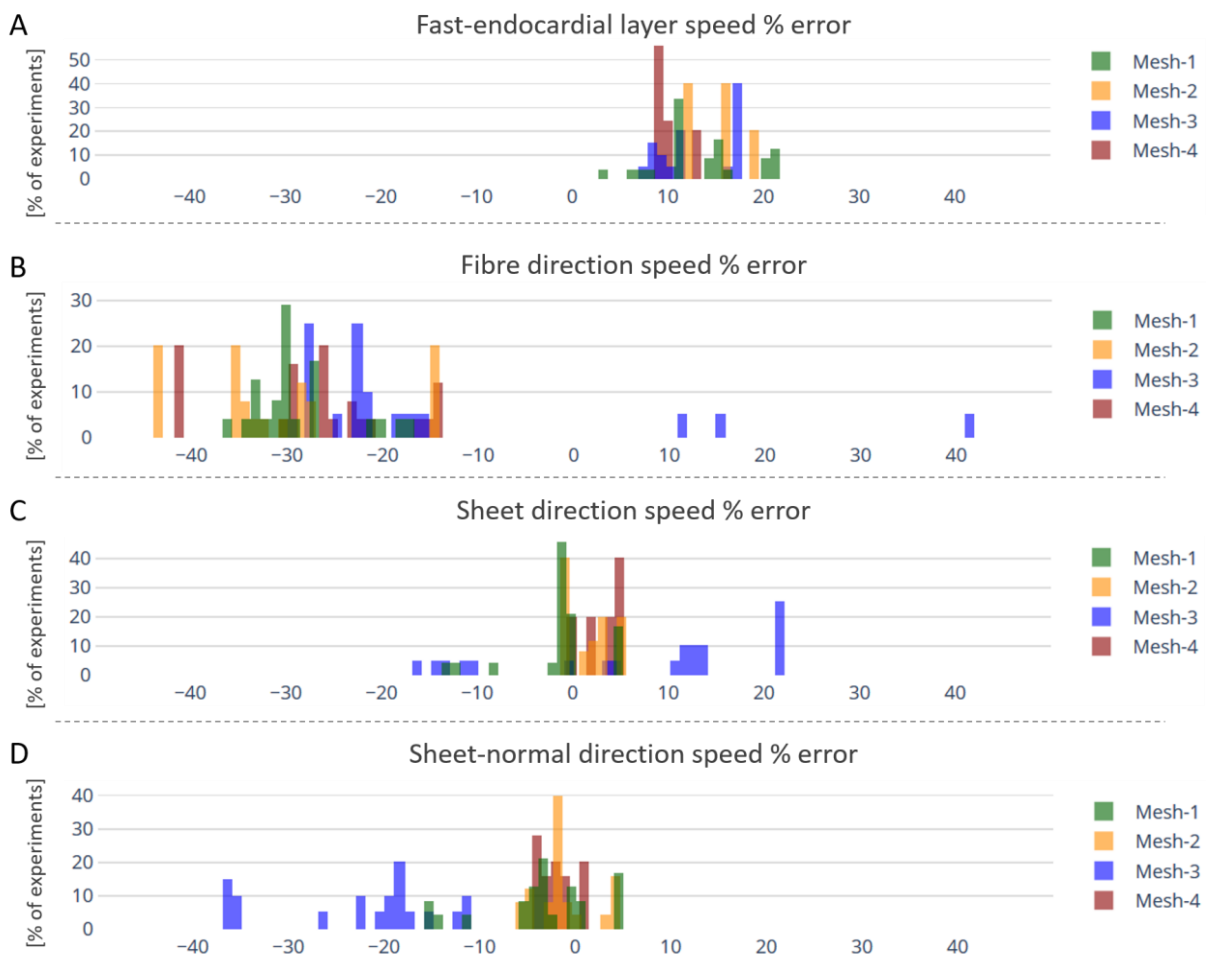

Fig. 5. Error in the conduction speeds inference from activation maps and low resolution for the root node discretisation. The figure shows the inference errors as the percentage over the ground truth conduction speeds (x-axis) represented as histograms grouping the results by anatomy. The y-axis represents the percentage of experiments with a given inference error from the 25 experiments per anatomy (5 speed-configurations 'times' 5 repetitions). Subfigures 5.A, 5.B, 5.C, and 5.D correspond to the errors from the endocardial, the fibre, the sheet, and the sheet-normal speeds, respectively.

Our inference process from epicardial activation maps reported inference errors (Section 2.7) whose absolute values were consistently lower than 25%, 45%, 20%, and 37% for the endocardial, fibre, sheet, and sheet-normal speeds, respectively. The median of these absolute error distributions was 11.7%, 28%, 3.1%, and 3.8%, respectively. In agreement with Fig. 4, Fig. 5 shows that the recovery errors for the fibre and sheet-normal speed were higher than for other speeds in most experiments. Overall, Fig. 5 demonstrates that the use of a low-resolution discretisation increased the speed inference errors compared to the high resolution (Fig. 4), especially in the experiments with Mesh-3.

## 3.4 Root node inference from ECGs

Analogously to Fig. 3, Fig. 6 illustrates the inferred root node locations when using 12-lead ECG-QRS as the target of the inference process for all virtual subjects with Mesh-4 (see Appendix A.6.2 for Mesh-1, Mesh-2, and Mesh-3).

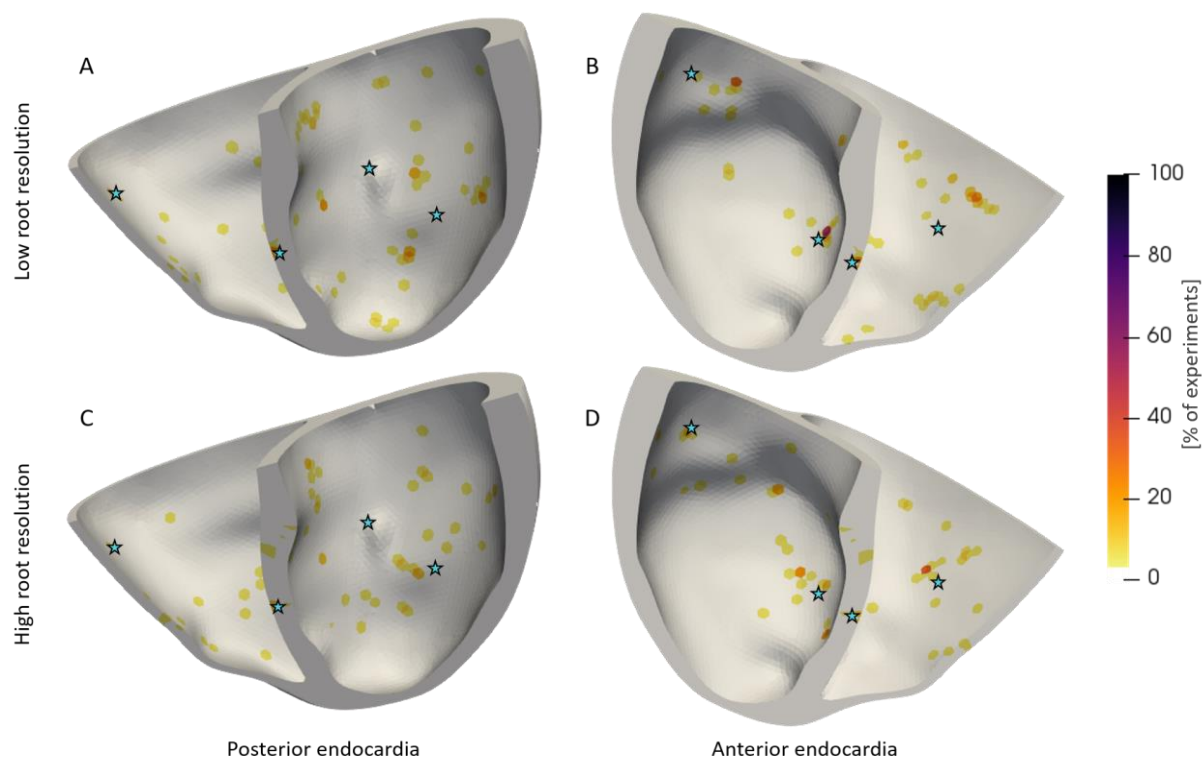

*Fig. 6. Root nodes inferred from 12-lead ECG on Mesh-4. This figure shows the root node locations inferred from the experiments with 12-lead ECGs as the target and with this torso-biventricular mesh. Subfigures 6.A and 6.B correspond to the predictions from using the low resolution for the root node discretisation; 6.C and 6.D correspond to the high resolution for the root node discretisation. The colourmap units are the percentage of the 25 inference experiments (5 repetitions 'times' 5 virtual subjects) that predicted a specific location. The star-icons denote the ground-truth locations.*

As illustrated in Fig. 6, as well as Fig. A.5, Fig. A.6, and Fig. A.7 in Appendix A.6.2, the recovery accuracy of the root node locations in the anterior endocardia was higher than in the posterior endocardia. The overall accuracy decreased compared to the results from epicardial activation map data (Fig. 3). Overall, it was unclear if the high resolution for the root node discretisation produced more accurate locations compared to the low-resolution strategy from ECG data.

## 3.5 Speeds inference from ECGs

Analogously to Fig. 4, Fig. 7 presents the inference percentage error for each speed from the experiments that used ECGs as the target and implemented the high resolution for the root node discretisation.

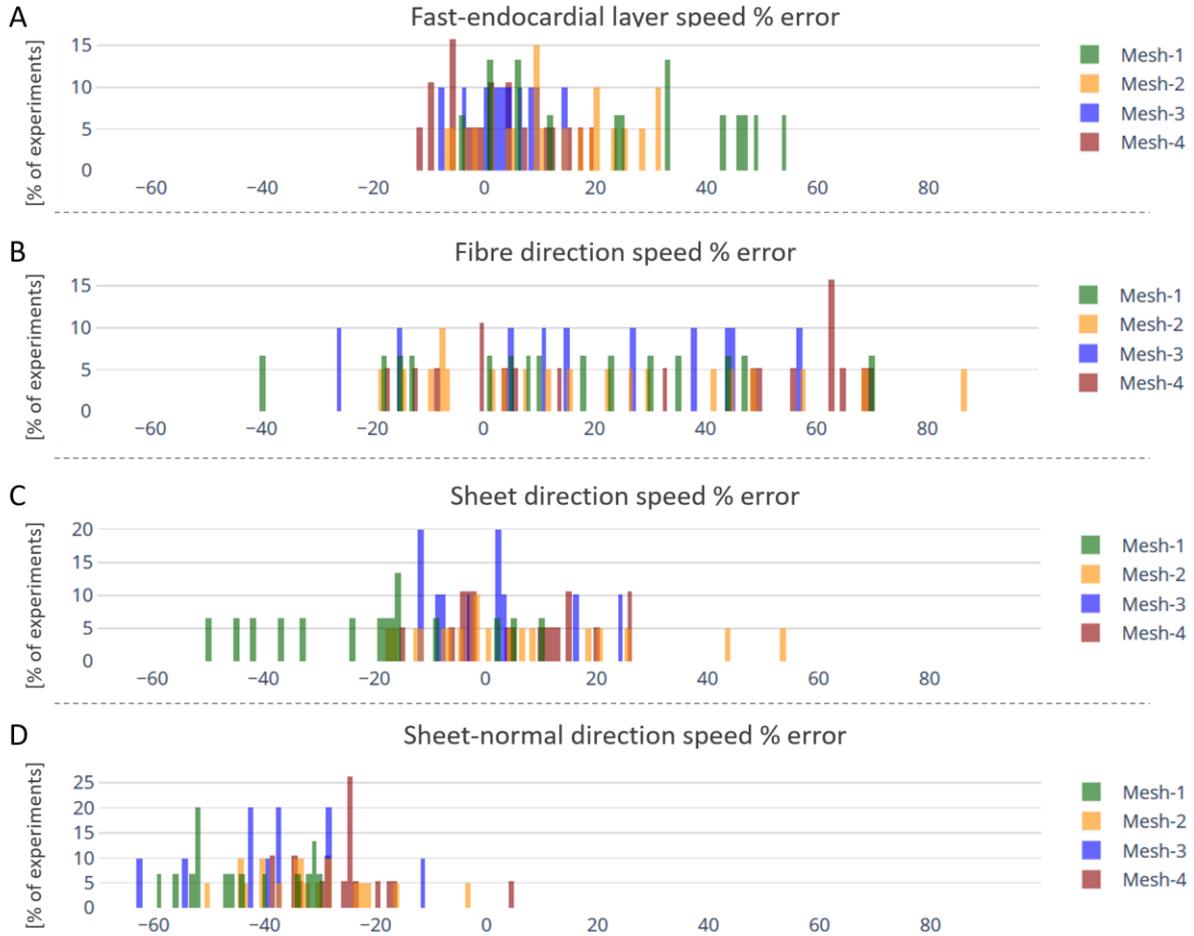

*Fig. 7. Error in the conduction speeds inference from ECGs and high resolution for the root node discretisation. The figure shows the inference errors as the percentage over the ground truth conduction speeds (x-axis) represented as histograms grouping the results by anatomy. The y-axis represents the percentage of experiments with a given inference error from the 25 experiments per anatomy (5 speed-configurations 'times' 5 repetitions). Subfigures 7.A, 7.B, 7.C, and 7.D correspond to the errors from the endocardial, the fibre, the sheet, and the sheet-normal speeds, respectively.*

Fig. 7 demonstrates that our approach reported inference errors (Section 2.7) with absolute values less than 30%, 70%, 25%, and 55% for the endocardial, fibre, sheet, and sheet-normal speeds, respectively, in most experiments featuring 12-lead ECG-QRSs as target and implementing the high resolution for the root node discretisation. The median of the absolute percentage errors were 9.4%, 22.9%, 11.2%, and 33.8%, respectively.

Fig. 7 also shows that the errors for the fibre speeds are entirely spread over the feasible range compared to other speeds, and all experiments underestimated the value of the sheet-normal speed. Moreover, the Mesh-1 (smallest mesh) accounted for the highest inference errors in endocardial and sheet speeds.

On the other hand, Fig. 8 presents the inference percentage error for each speed from the experiments that used ECGs as the target and implemented the low resolution of the root node discretisation.

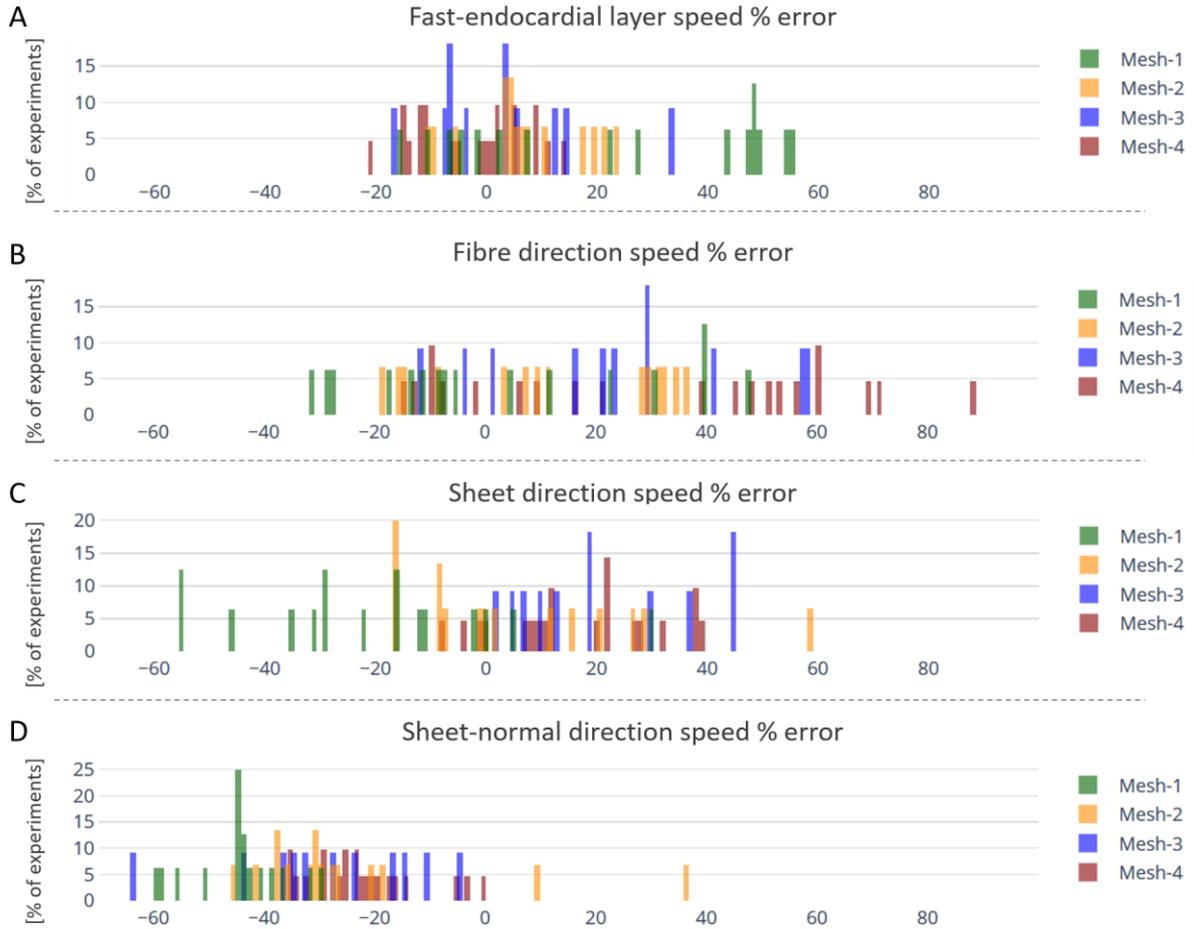

*Fig. 8. Error in the conduction speeds inference from ECGs and low resolution for the root node discretisation. The figure shows the inference errors as the percentage over the ground truth conduction speeds (x-axis) represented as histograms grouping the results by anatomy. The y-axis represents the percentage of experiments with a given inference error from the 25 experiments per anatomy (5 speed-configurations 'times' 5 repetitions). Subfigures 8.A, 8.B, 8.C, and 8.D correspond to the errors from the endocardial, the fibre, the sheet, and the sheet-normal speeds, respectively.*

Fig. 8 demonstrates that our approach reported inference errors (Section 2.7) with absolute values less than 30%, 60%, 40%, and 50% for the endocardial, fibre, sheet, and sheet-normal speeds, respectively, in most experiments featuring 12-lead ECG-QRSs as target and implementing the low resolution for the root node discretisation. The median of the absolute percentage inference errors were 9.5%, 22.3%, 16.3%, and 31.1%, respectively.

In agreement with Fig. 7, Fig. 8 also reports spread out error values for the fibre speed, underestimated sheet-normal speed values, and larger inference errors for Mesh-1 (smallest mesh) in the endocardial and sheet speeds.

## 4   Discussion

This study presents a novel efficient SMC-ABC-based inference method combined with Eikonal simulations using CMR-based torso-biventricular models for the estimation of the ventricular activation properties from the 12-lead ECG or epicardial activation time map data. We conducted the simultaneous inference of endocardial and myocardial conduction speeds, and the location and number of the root nodes in the endocardium, as these properties determine the activation sequence in the ventricles. The evaluation of the method was conducted on a cohort of twenty virtual non-diseased subjects to consider the effect of functional (five conduction speed configurations) and anatomical (four anatomies with biventricular volumes ranging from 74 cm$^3$ to 171 cm$^3$) variability in the healthy human population on the performance of the inference process. These experiments were conducted considering a high and a low resolution of root node discretisation to test the importance of the spatial precision in the algorithm's identification of these locations.

Our method was able to find populations of models that produced activation sequences with nearly identical patterns in activation maps and QRS complexes (width and morphology) than those observed from the subject's data. We observed that activation sequences mainly yield information about the root node locations, and the endocardial and sheet speeds. We also observed that, when using ECGs, the lowest estimation errors for endocardial and sheet speeds were given by the largest ventricular volumes. Moreover, from ECGs, the accuracy for the root node locations was higher for the root nodes located in the anterior versus posterior of the ventricles; whereas, from epicardial activation maps, the root nodes in the septum were the least accurately recovered.

A key contribution of this study is the development of an inference method to estimate the two types of parameters, namely, the anisotropic conduction speeds (continuous), and the number and locations of the root nodes (discrete) in both ventricles simultaneously. Previous work considered either known number of root nodes in simple cases or known number and location. For example, Giffard-Roisin et al. (2017) estimated the location of two root nodes in the left ventricle and the conduction speeds from ECG data; whereas, Grandits et al. (2020) estimated the activation times of a known set of root nodes and the conduction speeds from epicardial activation maps.

Our results demonstrate that different calibrations of activation properties can produce similar activation maps and ECGs to the data from a subject (Fig. 2) by inferring populations of parameter sets. The variability in the inferred population can serve to identify different possible sources to that target data. However, the precision at which a property can be recovered depends on its influence on the activation sequence pattern. Therefore, understanding the capabilities of the data and models to represent the effect of each activation property is crucial to identify the key factors that contribute to and affect the target data.

All our results for the inference of conduction speeds (Fig. 4, Fig. 5, Fig. 7, and Fig. 8) demonstrate that the endocardial and the sheet speeds were better identified than the fibre and sheet-normal ones in healthy ventricles in sinus rhythm. When assuming healthy conditions, defined as well-distributed root nodes and homogeneous tissue-conduction properties, the endocardial and the sheet-directed speeds (alongside the root nodes) dominate the activation sequence patterns. Therefore, the accurate recovery of the fibre and sheet-normal speeds became impossible in sinus rhythm. This phenomenon is due to their negligible impact on the resulting activation maps or ECGs, namely the impact of the fibre and sheet-normal speed was one order of magnitude less compared to the endocardial and sheet speeds. On the other hand, we conjecture that these speeds have a more substantial role in paced or pathological activation sequences, especially, if the conduction properties of the tissue are heterogeneous (e.g. fibrotic areas) or the root node locations are poorly distributed (e.g. bundle branch block). Consequently, focussing only on the endocardial and sheet-directed speeds when estimating the conduction speeds under healthy conditions can save in computational resources without impacting the accuracy of the results.

The endocardial and sheet speed estimation errors from ECGs were higher for the experiments with the smallest anatomies (Mesh-1 and Mesh-2) than for our other meshes (Fig. 7 and Fig. 8). This volumetric dependence of the error was caused by the calibration of our DTW-based discrepancy metric. As previously mentioned, our DTW algorithm measures two mismatch values in time and amplitude. Next, the DTW-based metric weights these two terms and adds them together, producing the discrepancy value. This discrepancy, then, informs the algorithm about 'how accurate' the evaluated particle is. This strategy was based on the notion that speeds account for time mismatches and root nodes for the morphological ones. However, the time mismatches depend on the conduction speeds and the volume of the mesh. Therefore, without rescaling for the mesh volume, the contribution from the time mismatch becomes negligible for small anatomies or overtakes the morphological mismatch component for large anatomies within the variability in the healthy human population. Based on this finding, we suggest that future studies aiming to estimate the conduction speeds from ECG recordings scale the trade-off between the temporal and morphological differences of these signals based on the physical characteristics of the subject.

Our root node inference results (Fig. 5, Fig. A.5, Fig. A.6, and Fig. A.7) from ECG data show that the anterior locations were better identified than the posterior ones. This asymmetric identifiability of the root nodes is a manifestation of the asymmetry of the electrode positioning protocol in the standard 12-lead ECG test. More precisely, each root node has a local effect on the activation sequence for its region on the heart that, later, gets represented by the ECG. However, the amplitude of the recorded regional-ECG is inversely proportional to the distance between that region and the electrode, while most electrodes are typically positioned on the left-anterior surface of the torso. Consequently, most changes in the patterns of the activation sequence in the posterior walls get masked by the electrical activity taking place in the anterior half of the organ for the precordial electrograms. We suggest future works to consider symmetrically placed electrodes around the torso to identify all root node locations from ECG data. On the other hand, when using epicardial activation maps to guide the inference, the root node recovery accuracy was better for root nodes in non-septal than in the septal areas since the root nodes in the septum have little influence on the epicardial activation time map.

Overall, we attained higher accuracies from using epicardial activation maps for the inference than from ECG recordings. The activation map is well-known to represent more information about the activation sequence compared to the QRS. However, the main advantage of the activation map is that the data are spatially distributed on a surface (epicardium) that strongly relates to the root node parameter space (except for the septal area). More precisely, the inference from epicardial activation maps can be subdivided into regional sub-problems since local root nodes only influence local patterns in the activation sequence. This phenomenon enables the algorithm to quickly identify partial solutions to the inference problem from epicardial activation maps. On the other hand, all root nodes influence the morphology of each lead in the 12-lead ECG, simultaneously. Thus, the ECG similarity can inform on how many root nodes are correct, but not exactly which.

From the experiments guided by activation maps, we observed that using a low-resolution discretisation, where the root nodes were far from each other, aided the algorithm to identify the number of root nodes (Fig. 2 and Appendix A.6.1). On the other hand, increasing the discretisation's resolution improved the resemblance of the predicted data (activation maps or ECGs) to the target (Fig. 2), as well as the accuracy of the inference for the conduction speeds and root node locations. However, these improvements were usually at the expense of overestimating the number of root nodes to surround the ground truth locations when these were unreachable due to the discretisation of the endocardial surface. Consequently, the k-means based root node combination strategy showed great synergy with the high-resolution discretisation, since k-means collapsed these additional locations into centroids that were closer to the ground truth sites than the nodes predetermined by the discretisation strategy.

Overall, the errors to recover the conduction speeds decreased with the discretisation's resolution of the root node parameter space (Fig. 4 and Fig. 5). In particular, when increasing the root node discretisation's resolution, the inference errors from Mesh-3 decreased from 17% to 7% and from 22% to 3% for the endocardial and sheet speeds, respectively. Moreover, the results from the ECG-based experiments agree with this finding by exhibiting a similar error increase when lowering the discretisation's resolution. These results suggest that the method's ability to infer the root nodes conditions its ability to estimate the conduction speeds.

As future work, we encourage that our methods are validated, for example, considering noisy data inputs as in Grandits et al. (2020), and, when possible, using clinical recordings, such as in Giffard-Roisin et al. (2017).

## 5 Acknowledgements

This work was funded by a Scatcherd European Scholarship, the Engineering and Physical Sciences Research Council, a Wellcome Trust Fellowship in Basic Biomedical Sciences to Blanca Rodriguez (214290/Z/18/Z), the CompBioMed 2 Centre of Excellence in Computational Biomedicine (European Commission Horizon 2020 research and innovation programme, grant agreement No. 823712), the Australian Research Council Centre of Excellence for Mathematical and Statistical Frontiers (CE140100049), and an Australian Research Council Discovery Project (DP200102101).

The computation costs of this work were incurred through an Amazon Web Services Machine Learning Research Award.

The authors would like to thank Dr Ernesto Zacur for generating the coarse biventricular-torso meshes utilised in this study.

# Appendix A

A.1. Bidomain model

The bidomain model is the gold standard formulation for cardiac electrophysiology simulations. The bidomain assumes that each point in space represents an average between multiple cells and that there are two electrical subspaces in the heart, namely, the intracellular space and the extracellular. The combination of both assumptions implies that at every point in space we have an intracellular and an extracellular potential, $\varphi_i$ and $\varphi_e$, respectively. Therefore, transmembrane potential, $V_m$, is defined as $V_m = \varphi_i - \varphi_e$. The current densities, $\bar{J}_i$ and $\bar{J}_e$, in these two spaces are defined as

$$\bar{J}_i = g_{ix}\frac{\partial \varphi_i}{\partial x}\bar{a}_x + g_{iy}\frac{\partial \varphi_i}{\partial y}\bar{a}_y + g_{iz}\frac{\partial \varphi_i}{\partial z}\bar{a}_z$$

$$\bar{J}_e = g_{ex}\frac{\partial \varphi_e}{\partial x}\bar{a}_x + g_{ey}\frac{\partial \varphi_e}{\partial y}\bar{a}_y + g_{ez}\frac{\partial \varphi_e}{\partial z}\bar{a}_z,$$

where $\bar{a}_x$, $\bar{a}_y$, and $\bar{a}_z$ are unit vectors along the $x$, $y$, and $z$ fibre-axes, respectively. More precisely, $x$ represents the fibre (same direction as the fibre), $y$ the sheet (perpendicular direction to the fibre, but along the same tissue sheet), and $z$ the sheet-normal (normal to the tissue sheet plane) fibre-directions. Each fibre-direction has an associated conductivity, $g$. Thus, $g_x$, $g_y$, and $g_z$ are the conductivities along for each of the fibre-directions. Moreover, these conductivities can vary between the intracellular and extracellular domains.

The bidomain model imposes the conservation-of-current constraint such that any current leaving one subspace must enter the other one. Thus, we can write $-\nabla \cdot \bar{J}_i = \nabla \cdot \bar{J}_e = I_v$, where $I_v$ is the transmembrane current per unit volume.

Ohm's law, $\bar{J} = \sigma\bar{E}$, defines a current density, $\bar{J}$, as the product of an electrical field, $\bar{E}$, and some conductivities, $\sigma$. Furthermore, assuming that the magnetic field is not time-varying, through Faraday's law, we can redefine the electrical field in terms of the electrical potential, $\bar{E} = -\nabla\varphi$. By applying both laws, we can rewrite the current densities as $\bar{J}_i = -\sigma_i\nabla\varphi_i$ and $\bar{J}_e = -\sigma_e\nabla\varphi_e$.

Next, by applying the conservation of currents assumption that relates the two domains through the cells' membranes, we can rewrite the density currents in terms of the transmembrane current, $\bar{I}_m$, as follows

$$\beta I_m = \nabla \cdot (\sigma_i \nabla \varphi_i)$$

$$\beta I_m = -\nabla \cdot (\sigma_e \nabla \varphi_e),$$

where $\beta$ accounts for the membrane surface to volume ratio.

Moreover, by employing the capacitance formulation, the transmembrane current can be written as $I_m = C_m\frac{\partial V_m}{\partial t} + I_{ion}(V_m, \eta)$, where $C_m$ is the cell's membrane capacitance, $I_{ion}$ is the ionic current that leaks in and out of the cell due to its mechanisms, and $\eta$ is the set of variables of the cell model. These ionic currents are governed by a set of non-linear differential equations that will vary for each cell model implementation.

Finally, the bidomain model is

$$\nabla \cdot (\sigma_i \nabla \varphi_i) = \beta(I_m = C_m \frac{\partial V_m}{\partial t} + I_{ion}(V_m, \eta))$$

$$\nabla \cdot (\sigma_e \nabla \varphi_e) = -\beta(I_m = C_m \frac{\partial V_m}{\partial t} + I_{ion}(V_m, \eta)).$$

## A.2. Electrical propagation Eikonal model

The bidomain equations provide a reaction-diffusion-based biologically detailed description of the electrical propagation in the human heart by assuming two electrical subspaces, namely the intracellular and the extracellular space. Computation of bidomain simulations is costly; thus, simplifications such as the Eikonal equation have been proposed to study the mechanisms defining the cardiac activation sequence (Colli Franzone, Guerri, and Rovida 1990; Wallman, Smith, and Rodriguez 2012; Cedilnik and Sermesant 2020).

The Eikonal model converts simulation of electrical wave propagation in a cardiac mesh into a shortest-path-finding problem. The Eikonal method produces equivalent activation sequences to bidomain or monodomain models for only a fraction (up to three orders of magnitude lower) of the computational cost (Wallman, Smith, and Rodriguez 2012). Moreover, the Eikonal method can simulate propagation on coarse meshes, whereas the bidomain approach has convergence restrictions on mesh resolution. For these reasons, we use the Eikonal model to generate activation map data in our inference method of the human ventricular activation properties.

The basic formulation of the Eikonal equation is $\sqrt{\nabla d^T \cdot \nabla d} = 1$, where $d$ is distance. However, we can rewrite the Eikonal equation to account for an electrical wavefront propagation time in an anisotropic cardiac mesh (Colli Franzone, Guerri, and Rovida 1990). Namely, we apply $\nabla d = v \nabla t$, where $v$ is speed in the fibre, sheet, and sheet-normal directions, $v(f, s, n)$, and $t$ is the traveling time $t(x, y, z)$ passing through a point $(x, y, z)$, so our formulation is $\sqrt{\nabla t^T \cdot V \cdot \nabla t} = 1$, where $V = \begin{bmatrix} x_l & x_t & x_n \\ y_l & y_t & y_n \\ z_l & z_t & z_n \end{bmatrix} \begin{bmatrix} v_f^2 & 0 & 0 \\ 0 & v_s^2 & 0 \\ 0 & 0 & v_n^2 \end{bmatrix} \begin{bmatrix} x_l & x_t & x_n \\ y_l & y_t & y_n \\ z_l & z_t & z_n \end{bmatrix}^T$, where $v_f$, $v_s$, and $v_n$ are the speeds in the fibre, sheet, and sheet-normal directed speeds, and where $[x_i, y_i, z_i]$ defines the vector $\vec{i}$, where $i$ can be either $f$, $s$, or $n$.

We implement the Eikonal equation by interpreting the tetrahedral-mesh as a connected graph where electric current can go directly from two connected nodes, $a$ and $b$, through an existing edge $(a, b)$, where the distance between $a$ and $b$ is the Euclidean distance, namely, $\|\overrightarrow{(a,b)}\|_2$. Therefore, from the Eikonal approach, we have that the time cost between two adjacent nodes $a$ and $b$ is $c_{a,b} = \sqrt{\overrightarrow{(a,b)}^T \cdot V^{-1} \cdot \overrightarrow{(a,b)}}$. Finally, we can solve the Eikonal's activation times using a multisource-multidestination extension of Dijkstra's algorithm (Dijkstra 1959). We set the root node locations as starting sites with $t = 0$ and all the other nodes of the biventricular mesh as destinations for which we want to calculate $t$.

## A.3. Virtual Subjects

This section provides further details on how the conduction speeds considered to design our cohort of 20 virtual subjects link to the conductivities employed by Mincholé et al. (2019).

We considered three fast-endocardial speeds to account for a slow (120 cm/s), normal (150 cm/s), and fast (179 cm/s) Purkinje network, and two myocardial speed scenarios, normal and fast. The normal myocardial conduction speeds (50, 32, and 29 cm/s for the fibre, sheet, and sheet-normal speeds, respectively) were chosen to match the [extracellular, intracellular] conductivity pairs used in the bidomain model, namely [5.46, 1.5], [2.03, 0.45], and [2.03, 0.225] millisiemens/centimetre (mS/cm) in the fibre, sheet and sheet-normal directions, respectively. Similarly, the fast myocardial conduction speeds (88, 49, and 45 cm/s) corresponded to conductivities [10.92, 3], [4.06, 0.9], and [4.06, 0.45] mS/cm. The root node locations were set to seven homologous root node locations (three in the right, and four in the left ventricle) following the findings from Cardone-Noott et al. (2016) on root node configurations that produced realistic healthy ECG recordings.

## A.4. Computation of the ECG from the activation time map

The QRS complexes in specific ECG leads were computed from the activation time maps simulated through the Eikonal model using the pseudo-ECG equation (Gima and Rudy 2002), as in Mincholé et al. (2019). For a given electrode location $(x', y', z')$, this equation takes the form

$$\Phi_e(x', y', z') = \frac{a^2 \sigma_i}{4\sigma_e} \int (-\nabla V_m) \cdot \left[\nabla \frac{1}{r}\right] dx\, dy\, dz,$$

where $V_m$ is the transmembrane potential, $\nabla V_m$ is its spatial gradient, $r$ is the Euclidean distance from a given point $(x, y, z)$ to the electrode's location, $a$ is a constant that depends on the fibre's radius, and $\sigma_i$ and $\sigma_e$ are the intracellular and extracellular conductivities, respectively. An ECG signal is obtained by then considering this integral throughout the ventricular activation sequence period.

Working from an activation map generated by the Eikonal approach, there is no cell model or time course data for $V_m$. However, a pseudo-ECG can still be produced (within a constant factor) by merely assigning $V_m$ to be one or zero depending on whether the node in question has been activated. The pseudo-ECG algorithm cannot recover the exact amplitude of the ECG leads, as does not take into account the different impedances and attenuations between each source $(x, y, z)$ and sensor $(x', y', z')$ points pair. Therefore, for this study, we always worked with standardised (i.e. mean zero and standard deviation one) ECG signals. This assumption allows us to ignore constants, and rewrite the pseudo-ECG equation as a sum over all mesh elements,

$$\Phi_e(x', y', z') = \sum_{j=1}^{N_{ele}} -(\nabla V_m)_j \cdot \left[\nabla \frac{s_j}{r_j}\right],$$

where $(\nabla V_m)_j$ is an estimated gradient over the $j$th tetrahedral element and $s_j$ its normalised volume. Distances $r_j$ are calculated using the centroids of each element. We estimate these gradients by assigning $V_{mi} = 1$ if the node $i$ is activated, and $V_{mi} = 0$ if it is not, which is sufficient to generate a scaled-amplitude ECG signal.

Fig. A.1 illustrates a comparison of the amplitude-standardised versions of the ECG recording obtained from using a diffusion model (blue) from the epicardium to the torso surface compared to the pseudo-ECG algorithm (green).

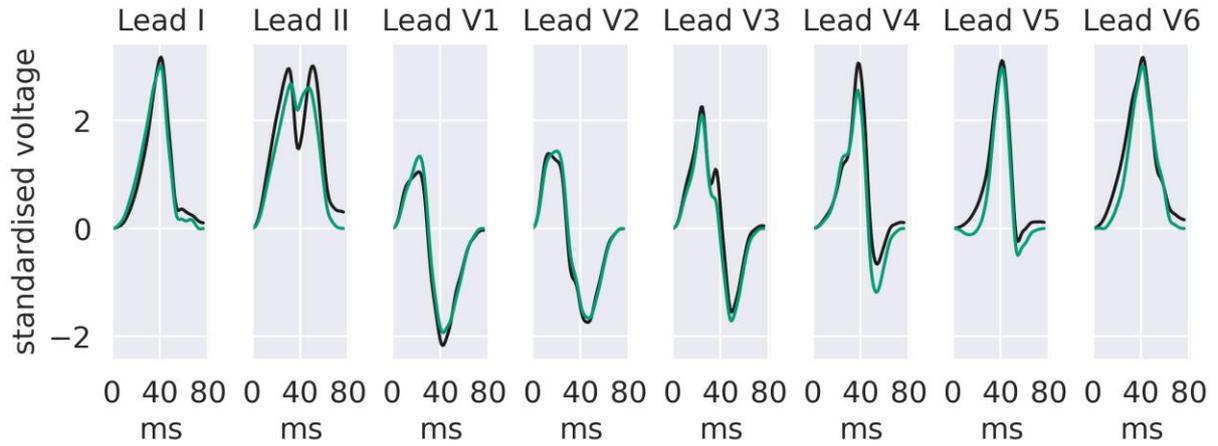

*Fig. A.1. ECG calculation justification example. ECG calculated as the diffusion from the bidomain-generated map to the electrodes (black). ECG calculated by the pseudo-ECG equation on the bidomain-generated activation time map (green). The amplitude of these signals is standardised to have mean zero and standard deviation one (y-axis). The time units are ms (x-axis).*

Fig. A.1 demonstrates that the pseudo-ECG equation produces realistic ECG recordings from the Eikonal model that can match the diffusion alternatives.

For the 12-lead ECG, we consider the eight independent leads. In other words, we ignore lead III and the augmented leads since they are linear combinations of other leads. Furthermore, simulating ECG signals from coarse meshes produces artefacts (Potse and Kuijpers 2010; Tate et al. 2019; Schuler et al. 2019). Thus, we filter the ECG recordings with a lowpass filter with a cutoff frequency set to 150 Hz following the Nyquist-Shannon sampling theorem and the guidelines in frequential information in the ECG from Sörnmo and Laguna (2005). Finally, the signals are aligned to start at 0 standardised voltage.

### A.5. SMC-ABC algorithm

Sequential Monte Carlo - approximate Bayesian computation (SMC-ABC) integrates the ideas behind SMC and ABC algorithms. SMC discovers regions of interest in the parameter space by gradually introducing the complexity of a given sampling problem. SMC evolves a population of particles through a series of intermediary distributions. Each mutation step employs the population to inform the subsequent sampling subproblem. This process results in a high degree of exploration initially, with each intermediate distribution trading some exploration for more exploitation in promising regions of the parameter space (similarly to simulated annealing [Lew et al. 2009]).

ABC targets an approximate posterior for model parameters $\theta$ (particle values), $p(\theta|\rho(d, d_{pred}) \leq \epsilon)$. Here $\rho(d, d_{pred})$ is a measure of discrepancy between the target data $d$, and $d_{pred}$, the predicted data from a particle $\theta$; and, $\epsilon$ is the cutoff discrepancy, with $\epsilon = 0$ recovering the true posterior. ABC defines a small discrepancy tolerance since the target data is sometimes impossible to reproduce due to noise.

As outlined in Step 5 from Fig. 1, the SMC approach consists of a resampling step that serves to replace low-quality particles (particles with high discrepancy), and then a mutation step that recovers particle uniqueness. As ABC does not define a likelihood, resampling is achieved simply by splitting the current particles into a 'to keep' and a 'to replace' group according to their discrepancy value and the current cutoff discrepancy. Then the algorithm replaces the 'to replace' group with an equivalent number of particles copied from the 'to keep' group, selected at random. The copied particles are then mutated by repeatedly implementing small changes that lower their discrepancies. Finally, the SMC-ABC accepts/rejects these proposed mutations according to the Metropolis-Hastings ratio, as in Markov chain Monte Carlo (Gilks 2005).

### A.6. Root node inference results

The main manuscript illustrates the root node inference accuracies for the experiments with Mesh-4 (torso and biventricular volumes of 44 dm$^3$ and 171 cm$^3$, respectively). This section reports the root node inference results for the remaining meshes, and the distance errors measured between the ground truth and the inferred locations. The discussion on these results can be found in the discussion section of the main manuscript.

#### A.6.1. Root node inference from epicardial activation maps

The mean ± standard deviation of the location error for all experiments with the low resolution of the root node discretisation and activation maps as the target was 1.07 ± 0.1 cm in the left ventricle, and 0.68 ± 0.15 cm in the right ventricle. The analogous error metrics for the inference of the number of root nodes were 0.65 ± 0.69 root nodes in the left ventricle, and 0.19 ± 0.24 root nodes, in the right ventricle. On the other hand, the inference errors from the high resolution of the root node discretisation were 0.48 ± 0.04 cm, and 1.68 ± 0.53 root nodes in the left ventricle, and 0.36 ± 0.06 cm, and 0.07 ± 0.17 root nodes in the right ventricle.

Fig. A.2 includes the root nodes inferred (after merging the populations of particles with k-means) from all experiments for the five virtual subjects with Mesh-1 with epicardial activation maps as the target. As mentioned in Section 2.1, Mesh-1 accounted for the smallest volumes in torso and ventricles (23 dm$^3$ and 74 cm$^3$, respectively).

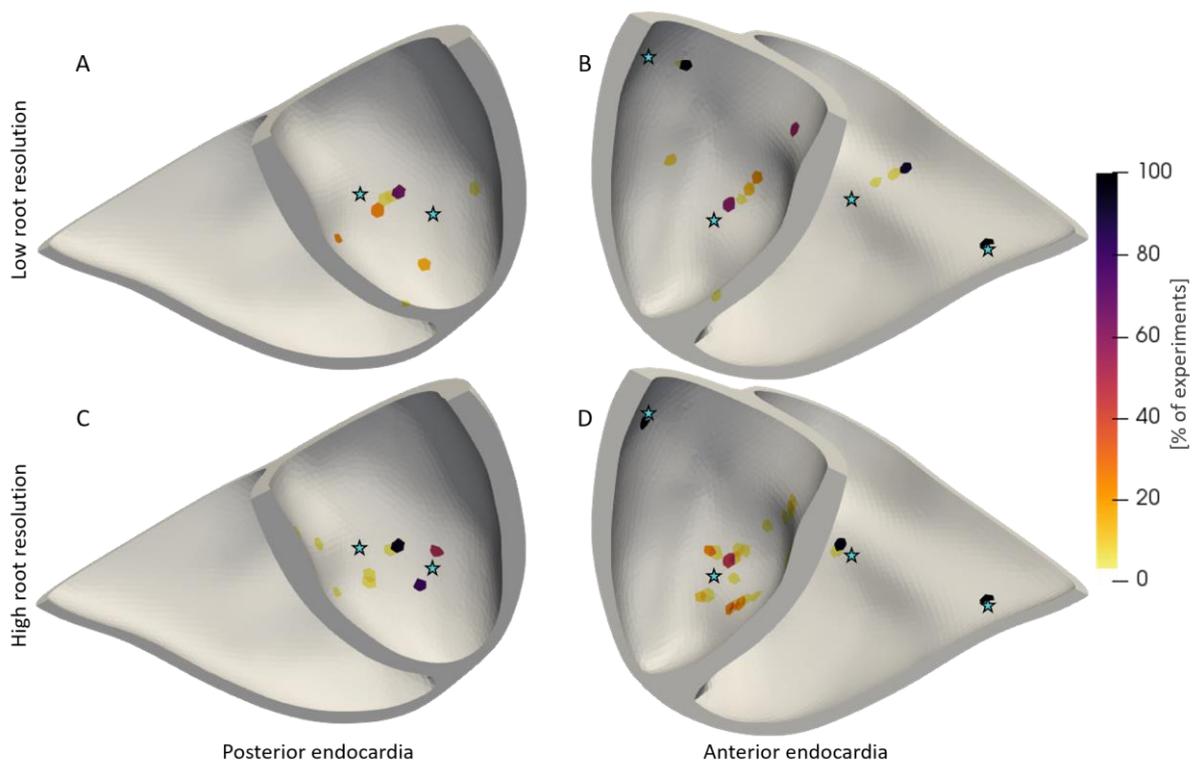

*Fig. A.2. Root nodes inferred from activation maps on Mesh-1. This figure shows the root node locations inferred from the experiments with epicardial activation time maps as the target and with this torso-biventricular mesh. Subfigures A.2.A and A.2.B correspond to the predictions from using the low resolution for the root node discretisation; A.2.C and A.2.D correspond to the high resolution for the root node discretisation. The colourmap units are the percentage of the 25 inference experiments (5 repetitions 'times' 5 virtual subjects) that predicted a specific location. The star-icons denote the ground-truth locations.*

Analogously to Fig. A.2, Fig. A.3 illustrates the root node inference results from the experiments with Mesh-2 (heart and torso volumes of 27 dm$^3$ and 76 cm$^3$, respectively).

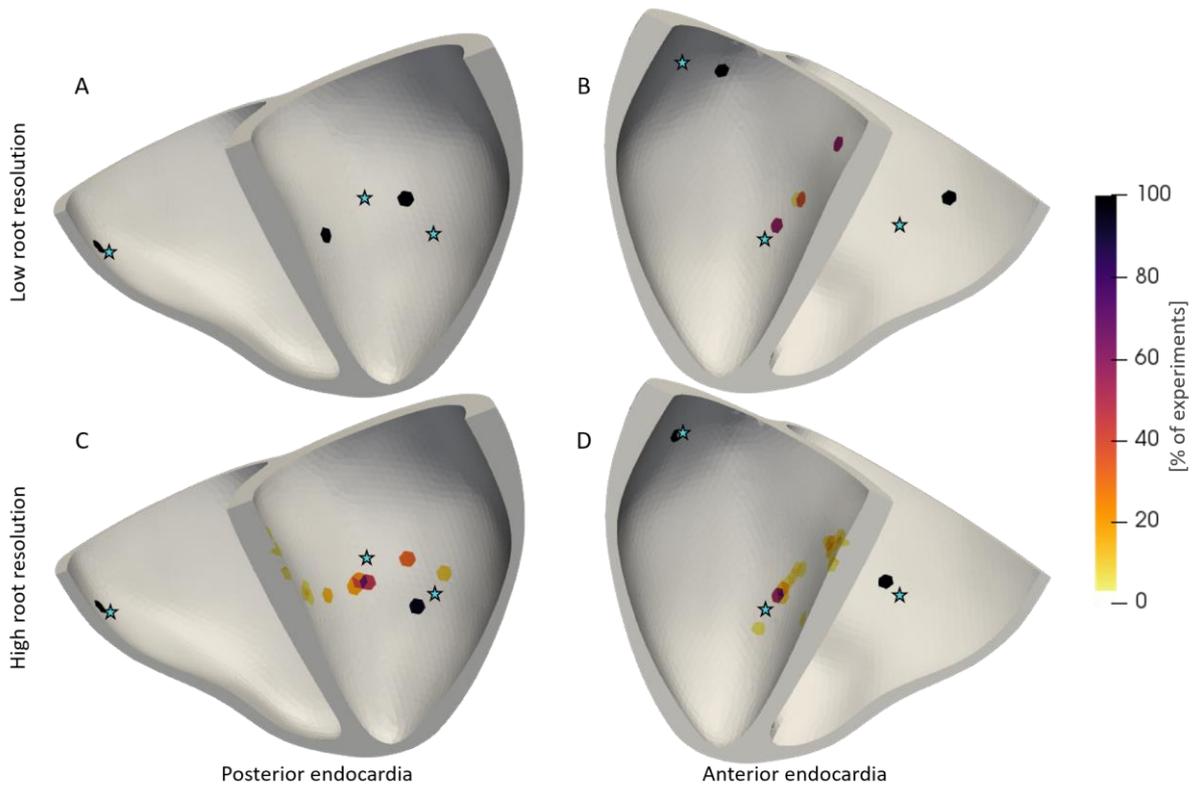

*Fig. A.3. Root nodes inferred from activation maps on Mesh-2. This figure shows the root node locations inferred from the experiments with epicardial activation time maps as the target and with this torso-biventricular mesh. Subfigures A.3.A and A.3.B correspond to the predictions from using the low resolution for the root node discretisation; A.3.C and A.3.D correspond to the high resolution for the root node discretisation. The colourmap units are the percentage of the 25 inference experiments (5 repetitions 'times' 5 virtual subjects) that predicted a specific location. The star-icons denote the ground-truth locations.*

Fig. A.4 accounts for the same results from Mesh-3 (heart and torso volumes of 54 dm$^3$ and 107 cm$^3$, respectively).

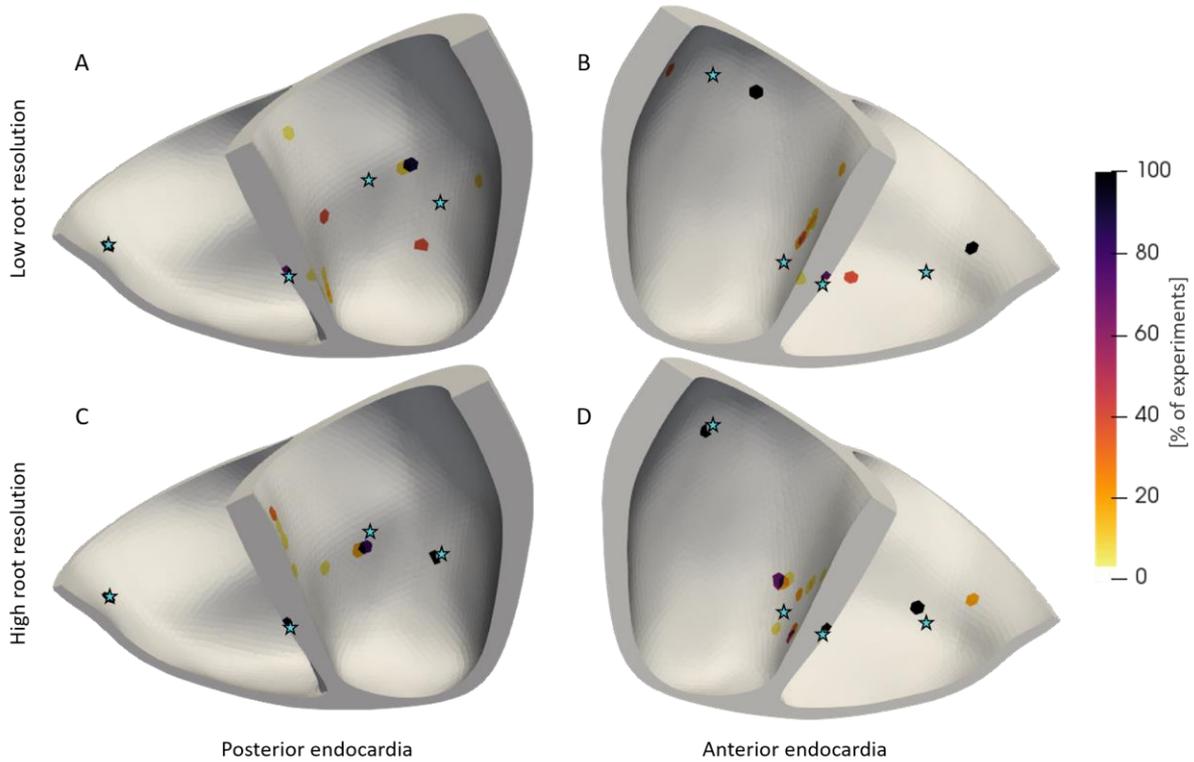

*Fig. A.4. Root nodes inferred from activation maps on Mesh-3. This figure shows the root node locations inferred from the experiments with epicardial activation time maps as the target and with this torso-biventricular mesh. Subfigures A.4.A and A.4.B correspond to the predictions from using the low resolution for the root node discretisation; A.4.C and A.4.D correspond to the high resolution for the root node discretisation. The colourmap units are the percentage of the 25 inference experiments (5 repetitions 'times' 5 virtual subjects) that predicted a specific location. The star-icons denote the ground-truth locations.*

Figures Fig. A.2, Fig. A.3, and Fig. A.4 show good agreement with the results from Fig. 3. Namely, less accurate predictions in the septal area, and more accurate predictions for the high than the low resolution of the root node discretisation.

### A.6.2. Root node inference from ECGs

Overall, the inference errors for all experiments with high resolution of the root node discretisation and epicardial activation map data in the left ventricles were 1.64 ± 0.46 cm (i.e. mean ± standard deviation), and 0.64 ± 0.86 root nodes, for the location and number of the root nodes, respectively. The errors in the right ventricles were 1.61 ± 0.56 cm, and 0.88 ± 0.87 root nodes, for the location and number of the root nodes, respectively. The experiments with the low resolution of the root node discretisation reported 1.82 ± 0.47 cm, and 0.72 ± 0.64 root nodes, for these same error metrics in the left ventricle; whereas, in the right ventricle we had 1.5 ± 0.47 cm, and 1.48 ± 1.23 root nodes, for the error in the location and number of the root nodes, respectively.

Similarly to the previous section, Fig. A.5, Fig. A.6, and Fig. A.7 illustrate the inferred root node locations from all experiments in the ECG domain from meshes Mesh-1, Mesh-2, and Mesh-3, respectively.

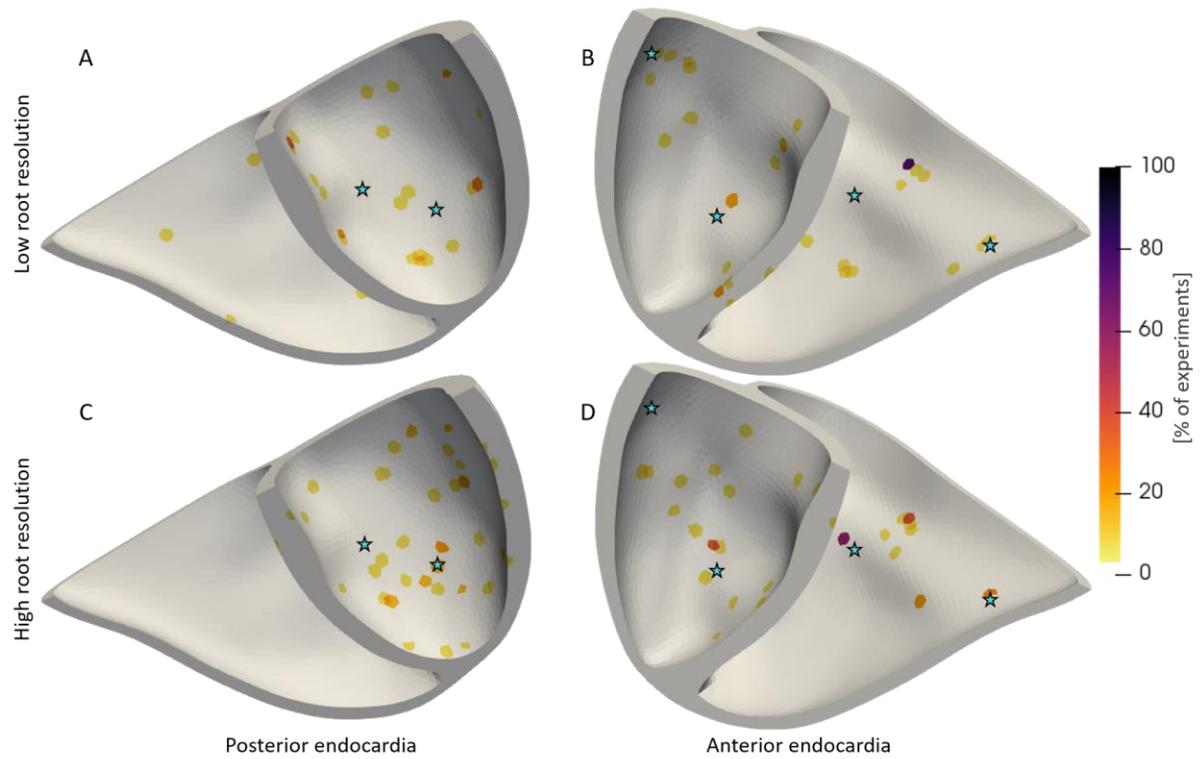

Fig. A.5. Root nodes inferred from 12-lead ECG on Mesh-1. This figure shows the root node locations inferred from the experiments with 12-lead ECGs as the target and with this torso-biventricular mesh. Subfigures A.5.A and A.5.B correspond to the predictions from using the low resolution for the root node discretisation; A.5.C and A.5.D correspond to the high resolution for the root node discretisation. The colourmap units are the percentage of the 25 inference experiments (5 repetitions 'times' 5 virtual subjects) that predicted a specific location. The star-icons denote the ground-truth locations.

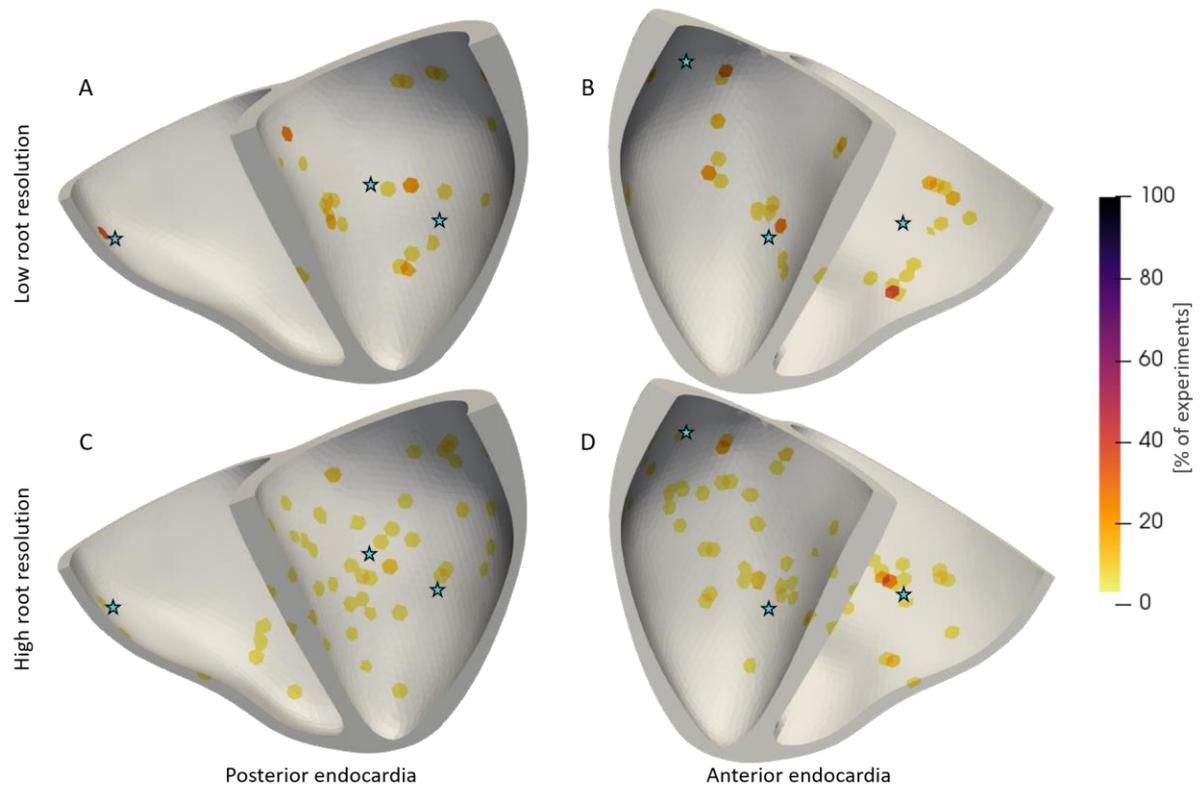

Fig. A.6. Root nodes inferred from 12-lead ECG on Mesh-2. This figure shows the root node locations inferred from the experiments with 12-lead ECGs as the target and with this torso-biventricular mesh. Subfigures A.6.A and A.6.B correspond to the predictions from using the low resolution for the root node discretisation; A.6.C and A.6.D correspond to the high resolution for the root node discretisation. The colourmap units are the percentage of the 25 inference experiments (5 repetitions 'times' 5 virtual subjects) that predicted a specific location. The star-icons denote the ground-truth locations.

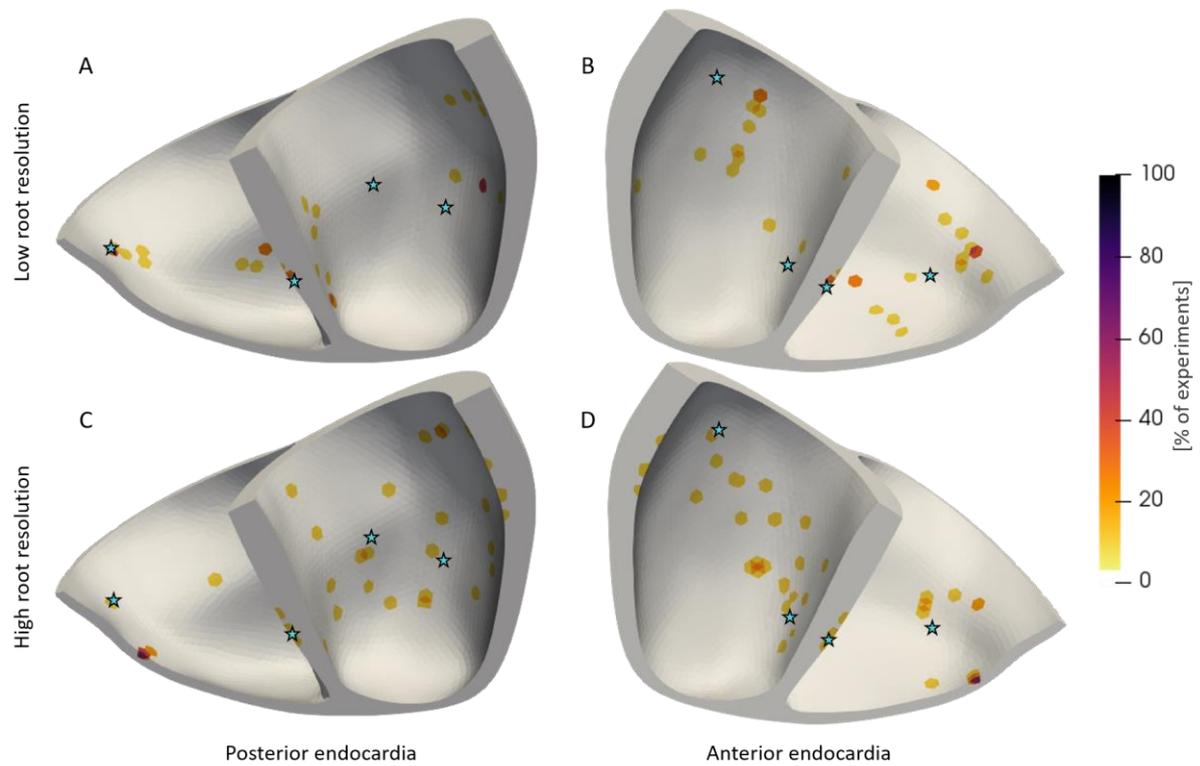

*Fig. A.7. Root nodes inferred from 12-lead ECG on Mesh-3. This figure shows the root node locations inferred from the experiments with 12-lead ECGs as the target and with this torso-biventricular mesh. Subfigures A.7.A and A.7.B correspond to the predictions from using the low resolution for the root node discretisation; A.7.C and A.7.D correspond to the high resolution for the root node discretisation. The colourmap units are the percentage of the 25 inference experiments (5 repetitions 'times' 5 virtual subjects) that predicted a specific location. The star-icons denote the ground-truth locations.*

Figures Fig. A.5, Fig. A.6, and Fig. A.7 show good agreement with the results from Fig. 6. Namely, more accurate predictions in the anterior half of the endocardia than the posterior half, and, overall, less accurate predictions than the results from the inference guided by epicardial activation maps (Fig. A.2, Fig. A.3, and Fig. A.4). Overall, the accuracy of the root node recovery was similar for all anatomies.